\documentclass{iau}
\usepackage{graphicx} 
\usepackage{natbib}
\newcommand{\mft}{m_{\rm FT}}

\newcommand{\snrft}{{\rm [S/N]_{FT}}}
\newcommand{\snrsp}{{\rm [S/N]_{SP}}}

\title[IAUS291.~~RRATs among pulsars] 
{Rotating Radio Transients and Their Place Among Pulsars} 

\author[S. Burke-Spolaor]  
{S. Burke-Spolaor$^1$
}

\affiliation{$^1$Jet Propulsion Laboratory, California Institute of Technology, 4800 Oak Grove Drive, Pasadena CA 91104 USA \\ email: {\tt sarah.burke-spolaor@jpl.nasa.gov} \\[\affilskip]}

\pubyear{2012}
\volume{291}  
\jname{\mbox{Neutron Stars and Pulsars: Challenges and Opportunities after 80 years}}
\editors{J. van Leeuwen, ed.} 
\begin{document}

\maketitle

\begin{abstract}
Six years ago, the discovery of Rotating Radio Transients (RRATs) marked what appeared to be a new type of sparsely-emitting pulsar. Since 2006, more than 70 of these objects have been discovered in single-pulse searches of archival and new surveys. With a continual inflow of new information about the RRAT population in the form of new discoveries, multi-frequency follow ups, coherent timing solutions, and pulse rate statistics, a view of RRATs' place in the pulsar population is beginning to form. Here we review the properties of neutron stars discovered through single pulse searches. We first seek to clarify the definition of the term RRAT, emphasising that ``the RRAT population'' encompasses several phenomenologies. A large subset of RRATs appear to represent the tail of an extended distribution of pulsar nulling fractions and activity cycles; these objects present several key open questions remaining in this field.
\keywords{pulsars: general, stars: neutron, stars: statistics}
\end{abstract}


\firstsection 
\section{The Discovery of and Interest in RRATs}
Pulsars were originally discovered through their single pulse emission \citep{psrdiscovery}. However, periodicity-based search techniques later dominated major pulsar search efforts due to the drastic gain in sensitivity they provided to typical pulsar signals \citep[e.\,g.][]{pksmb}. A recent renewal of interest in single pulse searches led \citet{rrats} to uncover sporadic pulses from eleven neutron stars whose signal was not detectable through periodicity searches in their 35-minute observations. These objects were dubbed Rotating Radio Transients (RRATs), and since 2006, single-pulse searches of archival and new pulsar surveys have revealed more than 60 further examples of such objects \citep[e.\,g.][]{deneva,keane,htruSP}.

If all RRATs' sporadically-detectable emission arises from nulling, the net Galactic population of these objects may be equal to or greater than that of steadily-emitting radio pulsars, generating a substantial discrepancy between neutron star birthrates and core-collapse supernova (CCSN) rates \citep{keanekramer}. This has given rise to questions about the intrinsic nature of RRATs, and how their observed behaviour may link to other pulsar populations or to a cycle in normal pulsar evolution. 
Motivated by this, below we make explicit what is meant by the term ``RRAT,'' and review what studies are revealing about these objects' relationship to other neutron star populations.


\section{Defining the RRAT}
\subsection{RRAT as a survey definition}
In recent papers reporting pulsar survey results, the most common definition of RRAT applies to a pulsar that was discovered in single pulses, but not in the periodicity-based search in the survey. The reasons that this may arise is: 1) A pulsar has a very high nulling fraction, thus a weak Fourier strength; 2) A pulsar has a pulse energy distribution with a weak average but extended tail; or 3) The presence of radio interference has caused a regularly-emitting pulsar's periodicity to experience a degradation of signal strength, or to be masked or passed over in the inspection process. Pulsars in the third category make up a minority ($\lesssim$5\%) of single pulse search results, and can generally be identified by excising interference from the discovery data or obtaining cleaner follow-up observations.

We can analytically approximate what level of nulling and modulated pulsars will lead pulsars in a survey to be considered RRATs. \cite{mclaughlincordes03} derived the signal-to-noise ratio found for a pulsar with unimodally and bimodally-distributed flux in a periodicity search and single pulse search. Let us consider a pulsar with single pulse above the survey detection threshold, where the average signal-to-noise ratio of detected pulses is given by $\langle\snrsp\rangle$. To be considered a RRAT, $\snrft$ must be less than the periodicity detection threshold, $\mft$. Following from McLaughlin \& Cordes, we derive:
\begin{equation}\label{eq:null}
{\rm NF} > 1-\frac{\eta}{\zeta}\cdot\frac{2\,\mft}{\langle\snrsp\rangle}\cdot\sqrt{\frac{P}{T_{\rm obs}}}~,
\end{equation}
where NF is the pulsar nulling fraction, $T_{\rm obs}$ gives the observation length, $P$ is the pulsar period, and $\eta$ and $\zeta$ are pulse-shape-dependent parameters of order $\sim$1. Similarly, a non-nulling, modulated pulsar will be considered a RRAT if:
\begin{equation}\label{eq:mod}
\frac{S_{\rm max}}{S_{\rm avg}} > \frac{\zeta}{\eta}\cdot\frac{\snrsp}{2\,\mft}\cdot\sqrt{\frac{T_{\rm obs}}{P}}~,
\end{equation}
where $S_{\rm max}$ is the maximum flux likely to be emitted over the observing interval (as given by the pulse energy distribution statistics), $S_{\rm avg}$ is the modified average pulse intensity defined by \citet{mclaughlincordes03}, and here $\snrsp$ represents the signal-to-noise ratio of the brightest pulse detected in the single pulse search.
What is immediately notable about these equations is that whether upon discovery a pulsar is called a ``RRAT'' is dependent on the pulsar's rotational period, and also highly survey dependent, changing with survey observing length and assumed detection thresholds (see also \citealt{keaneconfproc}).

\subsection{Discerning RRAT types}
Several methods exist to determine the underlying nature of RRATs. Perhaps the most effective of these is in pulse energy distribution analysis (e.\,g. as in \citealt{keane}, Miller et al.~in prep). This can uncover energy bimodality (implying a nulling pulsar) or reveal distributions in energy and time that are consistent with the lognormal or power-law distributions typically attributed to pulsars. Thus far, few analyses of such kind have been published, and so we cannot reliably assess the percentile breakdown of the underlying emission type of the $\sim$70 currently-identified RRATs. However, the cursory studies of \citet{bsb} and deeper studies of \citet{keane} have indicated that it is likely a large fraction of RRATs are in fact undergoing nulling (or, at least, are bimodally distributed with a low state below the detection thresholds).


\section{The Extreme Nature of RRATs in the Pulsar Population}
We now seek to quantify how excessive modulation and nulling are in RRATs when compared with the distribution of these phenomena in the general pulsar population.

\subsection{Pulse-to-pulse modulation}\label{sec:mod}
Previous studies have indicated that pulse-to-pulse energy density variations are typically lognormally distributed \citep[e.\,g.][]{cognard,cjd04}, or follow a power-law distribution (as with giant micropulses; \citealt{jr02}). Soon after the first report of RRATs, \citet{patrick} revealed that PSR\,B0656+14 has an extremely extended lognormal tail (such that $S_{\rm max}/S_{\rm avg}\simeq450$); they noted that were the pulsar at a greater distance, it would have been detected as a RRAT (fulfilling Eq.\,\ref{eq:mod} above). Recently, the first targeted measurements of single pulse energy distributions for a large pulsar sample were reported by \citet{htruV}. Their analysis suggested that long-tailed pulsars like B0656+14 are uncommon, and that most pulsars appear to cluster around a relatively narrow range in lognormal shape parameters.

\begin{figure}
\begin{center}
\includegraphics[width=0.75\textwidth]{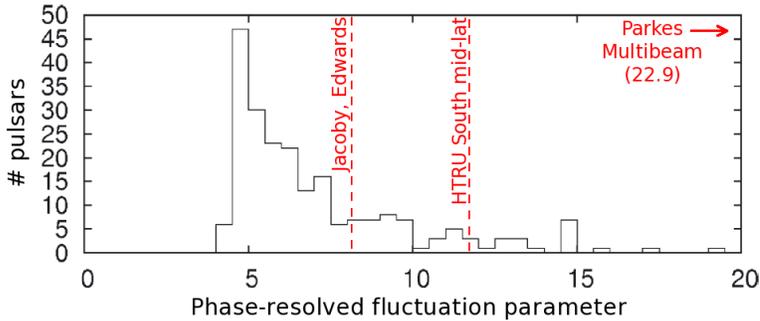} 
\caption{Here we show the distribution of the phase-resolved fluctuation parameter (here equivalent to $S_{\rm max}/S_{\rm avg}$) for a sample of 314 pulsars. We indicate the line above which a pulsar might appear as a RRAT in various surveys. See Sec.\,\ref{sec:mod} for discussion. Surveys are: \citet{jacoby}, \citet{edwards}; HTRU South mid-lat, \citet{htruSP}; Parkes Multibeam, \citet{pksmb}. This figure was adapted from \citet{htruV}.}
   \label{fig:mod}
\end{center}
\end{figure}

In Figure\,\ref{fig:mod}, we reproduce the distribution of phase-dependent $S_{\rm max}/S_{\rm avg}$ measured for that sample. Using Eq.\,\ref{eq:mod} and $\mft=6$ we have marked, for various surveys, the limit above which a pulsar with $P=1$\,s and $\snrsp=6$ will be discovered as a RRAT.\footnote{Note that this comparison is most accurate for surveys with a $T_{\rm obs}$ similar to that of the distribution's sample, 9\,minutes; if the distribution were made from shorter/longer observations, we would expect the distribution to extend to lower/higher values, respectively.} This appears to suggest that while single pulse searches may certainly uncover extremely modulated pulsars, pulsars do not necessarily have to exhibit excessive modulation to be found as RRATs in a search, particularly for surveys of relatively short duration.
 

\subsection{Pulse Nulling}
\begin{figure}
\begin{center}
 \includegraphics[width=0.98\textwidth]{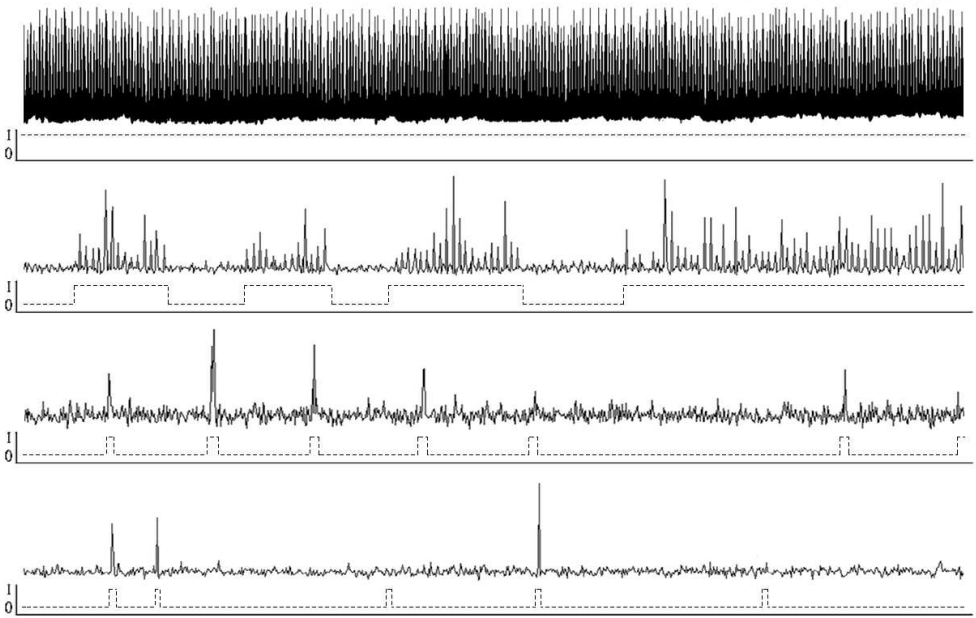} 
 \caption{Here we show time series for pulsars exhibiting a range of emission activity timescales (top to bottom: Vela, PSRs\,J1646--6831, J1647--36, J1226--32; archival data from \citealt{edwards}, all panels are of equal duration). Each time series has been dedispersed using the pulsar dispersion measure reported by the online {\sc psrcat} database \citep{psrcat}. The binary scale below each time series shows an estimated representation of the null/emission state. PSRs J1226--32 and J1647--36 were reported as RRATs by \citet{bsb}. PSR\,J1647--36 exhibits clusters of $\sim$5--10 pulses per activity cycle, while PSR\,J1226--32 emits singular pulses, perhaps signifying a typical emission timescale of less than the rotational period.}
   \label{fig1}
\end{center}
\end{figure}

There are three variations of nulling activity that have been recognised:
\begin{itemize}
\item ``Standard'' nulling \citep[e.\,g.][]{r76,wang}, in which nulling fractions range from a few percent up to $\sim$95\%, and activity timescales range seconds to minutes.
\item High-fraction nulling, i.\,e. as in some RRATs. Analysis of these objects indicates NFs upwards of 99\%. In objects discovered as RRATs, the typical null timescale far exceeds that of the emitting timescale (see Fig.\,\ref{fig1}).
\item Intermittent pulsars \citep{kramer06,fernandoint}, which undergo nulling and emission cycles on timescales of weeks to years. 
\end{itemize}
The activity timescales in both null and active states appear to follow characteristic spans for all of these variations, and some nulling appears quasi-periodic. The pulsation rate of most RRAT discoveries tend to be reproducible between observations, and they may also exhibit long-timescale quasi-periodicities \citep{pall}. 

This consistency allows a comparison of average emission/null cycle for the various manifestations of nulling. Such an analysis was performed by \citet{htruSP}, who found that the cycle times of \citet{wang} nulling pulsars appear continuously distributed with those of RRATs (as we would expect given Eq.\,\ref{eq:null}). The ${\rm NF}\gtrsim95$\% pulsars were exclusively highlighted by RRATs, with the highest exceeding $99.99$\%. Intermittent pulsars yet remain isolated from short-duration null cycle pulsars. This perhaps marks two distinct populations, however the lack of current surveys' sensitivity to intermediate-timescale null cycles indicates that perhaps there is yet an undiscovered population of pulsars undergoing such nulling.
It is the nature of extreme nulling pulsars (${\rm NF}>95$\%) which most motivates continued observations of RRAT discoveries, and on which we focus the remainder of this manuscript.

\section{Nulling Pulsars and the Issue of Birthrates}
It is not yet known what triggers nulling activity in pulsars, and whether there is some internal plasma/energy timescale intrinsic to the pulsar, or an external influence like asteroid bombardment \citep{lighthouse}. We have previously noted the issue of matching neutron star birthrates with the occurrence rate of the CCSN thought to produce pulsars; in this discourse we may find circumstantial clues to how nulling phenomena relate to other pulsar populations.

Taking the original 11 RRATs to be nulling objects, \citet{rrats} calculated the implied population of Galactic pulsars to be 2--$4\times10^5$, several times larger than the estimates for standard pulsars \citep{lorimer06}. The high-nulling-fraction population makes the largest contribution to the factor of 5--6 excess in neutron star birthrates, with other contributions made by radio quiescent populations such as X-ray dim isolated neutron stars, magnetars, and central compact objects \citep{keanekramer}. While it is perhaps feasible that birthrate estimates have been grossly overestimated, the CCSN rate has been grossly underestimated, or there is a highly effective alternate source of pulsar production, a very plausible solution to the birthrate discrepancy can be found by drawing an evolutionary link between nulling behaviours and other neutron star varieties.

\section{Extreme Nulling Pulsars: Evolutionary Clues from Recent Studies}
\subsection{Spin parameters of RRATs with high nulling fraction}
It has long been debated as to whether nulling fraction correlates with any pulsar spin parameters, particularly with $P, \dot P,$ or characteristic age ($\tau_c$), as might be the case if nulling is related to pulsar evolution. No definitive answer has yet been provided, particularly as nulling studies seem to disagree: \citet{r76} first reported a \mbox{${\rm NF}\propto\tau_c^{-1}$} relationship, whereas \citet{rankin} reported no clear correlation between these two parameters. \citet{biggs} subsequently reported only a weak NF--$\tau_c$ anti-correlation, and that in fact a stronger anti-correlation exists between the NF and the magnetic/rotational alignment of the pulsar beam. Most recently, \citet{wang} performed a dedicated study of nulling for pulsars in the Parkes Multibeam survey, again reporting a weak anti-correlation of age with NF, noting that there is a higher tendency for pulsars with a high NF ($\gtrsim$60\%) to be older than 5\,Myr.

Single pulse searches are intrinsically biased to discover nulling pulsars with longer periods (Eq.\,\ref{eq:null}), so a separate analysis of only RRATs cannot wholly inform the above discourse. However, ongoing timing observations \citep{mauratiming,evantiming} have shown that RRATs discoveries seem to support the Wang et al.~observation (most reported solutions show $\tau_c>$ a few Myr). Strikingly, the bridge in $P$--$\dot P$ space between canonical pulsars and both magnetars and X-ray dim isolated neutron stars is preferentially highlighted by RRAT discoveries. This seems to give a cursory indication that extreme nulling pulsars may somehow link these populations.


\subsection{PSR\,J1819--1458: Tying Extreme Nulling and Magnetars?}
One of the original \citet{rrats} discoveries, PSR\,J1819--1458, shows the strongest evidence for an evolutionary tie between neutron star populations. Its estimated surface magnetic field strength, $B_{\rm surf}=5\times10^{13}$\,G, is the highest known among RRATs, and lies just below the lowest known magnetar $B_{\rm surf}$ \citep{rrats}. Dedicated monitoring and timing of this pulsar led \citet{lyne} to report the occurrence of an ``anomalous glitch'' in the pulsar, where the post-glitch recovery led to a net decrease in $\dot P$, rather than the increase typically observed in glitching pulsars. The implication of repeated occurrence of such glitches is that PSR\,J1819--1458 would experience secular migration from magnetar-like spin parameters to those of standard radio pulsars. That PSR\,J1819--1458 is currently magnetar-like is also supported by the properties of its detection the X-ray and infrared wavebands \citep{1819xray,reaIR}.

Of course, PSR\,J1819--1458 is only one example of an extreme nulling pulsar, and we cannot presume it represents the entirety of this population. Ongoing studies will
reveal whether other objects reflect similar behaviours. Thus far only a few RRATs have been targeted for detailed study; \citet{mauratiming} has reported on six more of the original 11 RRATs, noting that no glitches have yet been observed in these pulsars. \citet{reaIR} furthermore found no infrared detection in PSR\,J1317-5759, and \citet{kaplan} put limits on X-rays from PSRs\,J0847--4316 and J1846--0257. 

We can identify several points of future study that will advance our understanding of RRATs. In particular, targeted differentiation of modulated/nulling RRATs, and the obtainment of timing solutions for these objects, will provide measured spin parameters, a large number of pulse detections, and precise positions for these objects. This will leading to NF studies that include extreme-nulling objects to search for correlations with other neutron star properties. Further detection of (or limits on) the presence of glitches in these objects will explore their spin evolution in comparison with canonical pulsars and other neutron star populations. Finally, obtaining precise RRAT positions will overcome issues of crowded high-energy fields, and may lead to further understanding of these objects through X-ray and other high-energy detection.

\section{Acknowledgements}
\noindent A portion of this research was carried out at the Jet Propulsion Laboratory, California Institute of Technology, under contract with the National Aeronautics and Space Administration.


\end{document}